\documentclass[aps,prb,superscriptaddress,notitlepage,twocolumn,10pt]{revtex4-1}

\usepackage[dvips]{graphicx}
\usepackage{array}

\usepackage{dsfont}
\usepackage{amsmath}
\usepackage{amssymb}
\usepackage{enumerate}

\usepackage[percent]{overpic}

\usepackage{color}

\def\bdm{\begin{displaymath}} \def\edm{\end{displaymath}}
\def\nn{\nonumber} \def\bc{\begin{center}} \def\ec{\end{center}}
\def\be{\begin{equation}} \def\ee{\end{equation}}

\def\vd{{\bf d}}   
   
\def\vg{{\mbox{\boldmath$g$}}}   
   
\def\vk{{\bf k}}   
   
\def\vn{{\bf n}}   \def\vone{{\bf 1}}

  \def\v0{{\bf 0}} 
  \def\kB{k_{\rm B}}
\def\kF{k_{\rm F}} \def\EF{E_{\rm F}}

\def\vDelta{{\mbox{\boldmath$\Delta$}}}

\def\vtau{{\mbox{\boldmath$\tau$}}}

\begin{document}


\title{Josephson effect and triplet--singlet ratio of non--centrosymmetric superconductors}


\author{Ludwig Klam}
\email[]{lklam@itp.phys.ethz.ch}
\affiliation{ETH Z\"{u}rich, Institut f\"{u}r Theoretische Physik, Wolfgang-Pauli-Str. 27, CH-8057 Z\"{u}rich}
\author{Anthony Epp}
\affiliation{Max-Planck-Institut f\"{u}r Festk\"{o}rperforschung, Heisenbergstrasse 1, D-70569 Stuttgart}
\affiliation{University of British Columbia, 6224 Agricultural Road, Vancouver, B.C., Canada, V6T 1Z1}
\author{Wei Chen}
\affiliation{Max-Planck-Institut f\"{u}r Festk\"{o}rperforschung, Heisenbergstrasse 1, D-70569 Stuttgart}
\author{Manfred Sigrist}
\affiliation{ETH Z\"{u}rich, Institut f\"{u}r Theoretische Physik, Wolfgang-Pauli-Str. 27, CH-8057 Z\"{u}rich}
\author{Dirk Manske}
\affiliation{Max-Planck-Institut f\"{u}r Festk\"{o}rperforschung, Heisenbergstrasse 1, D-70569 Stuttgart}


\date{\today}

\begin{abstract} 
We calculate the Andreev bound states and the corresponding Josephson current for an asymmetric 2--dimensional Josephson junction by solving Bogoliubov--de--Gennes equations. The junction consists of a non--centrosymmetric superconductor (NCS) separated by a tunneling barrier with a variable height to a conventional $s$--wave superconductor. In addition to the antisymmetric spin--orbit coupling in the NCS on the one side, this asymmetric junction gives rise to a Rashba spin--orbit coupling at the interface. We explore the rich parameter space and recover various limiting cases such as $s$--wave/$p$--wave junction and the asymmetric s--wave junctions. In addition, we report a transition from a $0$--junction to a $\pi/2$--junction with increasing triplet--singlet pairing ratio of the NCS, which serves as a novel mechanism to determine the unknown ratio in a variety of NCS's. 


\end{abstract}


\maketitle



%



\section{Introduction}
Non--centrosymmetric superconductors (NCS) provide the unique possibility to study a microscopic coexistence of spin singlet and spin triplet superconductivity in a bulk material~\cite{Gorkov:2001:01,Kimura:2005:01,Sugitani:2006:01,Klam:2012:01}. This is due to the absence of inversion symmetry which would allow to distinguish even (spin singlet) and odd (spin triplet) parity pairing states. In this context the presence of strong antisymmetric spin--orbit coupling (ASOC) plays an essential role. The parity--mixing of the pairing state is determined by the strength of ASOC and, in a much stronger way, by the pairing interaction. Interesting effects are expected in NCS if the two parities appear in comparable magnitude. One possible realization is CePt$_3$Si and several other Ce--based NSC, where magnetic fluctuations may mediate a sizable or even dominant odd--parity component~\cite{Bauer:2004:01,Klam:2012:01}. 

The question of how to determine the relative magnitude of the two parity components has been addressed in various ways.
B\o{}rkje and Sudb\o{}~\cite{Borkje:2006:01} proposed to consider steps in the current--voltage characteristics of NCS--NCS junctions. The observation of a zero--bias anomaly in quasiparticle tunneling spectroscopy of certain directions would indicate the presence of helical edge state of topologically non--trivial phase with dominant odd--parity pairing \cite{Iniotakis:2007:01}. As discussed first by  Vorontsov {\it et al.}~\cite{Vorontsov:2008:01} the helical states would carry spin currents which would be another indication for the topologically non--trivial state. Crossed Andreev reflection between two interfaces between spin--polarized normal metals and a NCS has been proposed as a further diagnostic tool by Fujimoto~\cite{Fujimoto:2009:01}.
Asano and Yamano~\cite{Asano:2011:01} considered a NCS--NCS junction and predicted a difference in the temperature dependence of the critical current giving insights into parity--mixing. 
Klam and collaborators suggested Raman scattering as a way to determine the odd--even parity ratios~\cite{Klam:2008:02}. Experimentally, Yuan {\it et al.}~\cite{Yuan:2006:01}  analyzed the temperature--dependent penetration depth for non--centrosymmetric Li$_2$Pd$_x$Pt$_{3-x}$B, as nodes can appear due to parity mixing. 

In this study we analyze yet another method to get insight into the ratio of even and odd parity component, based on the current--phase relation of the Josephson contact between a NCS and a conventional $s$--wave (BCS) superconductor. Indeed working Josephson contacts of this kind have been fabricated for CePt$_3$Si~\cite{Sumiyama:2005:01} and isostructural LaPt$_3$Si~\cite{Aoki:2010:01}, both coupled to Al. Hayashi et al.~\cite{Hayashi:2008:01} pointed out that the observed qualitative difference in the interference effect in a magnetic field for Al--CePt$_3$Si Josephson contacts could be understood in terms of 
selection rules for the even and odd parity components of the superconducting pairing state. While the selection rules are concerned with lowest order Josephson tunneling, we would like to extend our discussion including higher order couplings, restricting ourselves, however, to contributions of the Andreev bound states at the two--dimensional interface. These are giving the most relevant contributions to the deviations from ordinary current phase relations. We will here concentrate on systems like 
CePt$_3$Si and LaPt$_3$Si which have a tetragonal crystal symmetry and an ASOC with Rashba--like structure (point group C$_{4v}$), as we will introduce below.
In order to discriminate between different triplet--singlet ratios, the considered junction is oriented parallel to the c--axis (four--fold axis of the crystal) of the NCS. For that case, the triplet component of the gap, which is in the simplest case of $p$--wave type, will contribute at most to the Josephson current.

This paper is organized as follows. In Sec.~\ref{model} we present the model Hamiltonian and the ansatz to solve the Bogoliubov--de--Gennes equations. Sec.~\ref{analytical:results} reports analytical results for the Andreev bound states for certain limiting cases, and in Sec.~\ref{numerical:results} we present the current--phase relations for Josephson junctions with different values for the singlet and triplet order parameter.
Finally, we summarise our results in Sec.~\ref{summary}.

\section{Model\label{model}}

\begin{figure}[htbp]
\includegraphics[angle=0, width=1.0\linewidth]{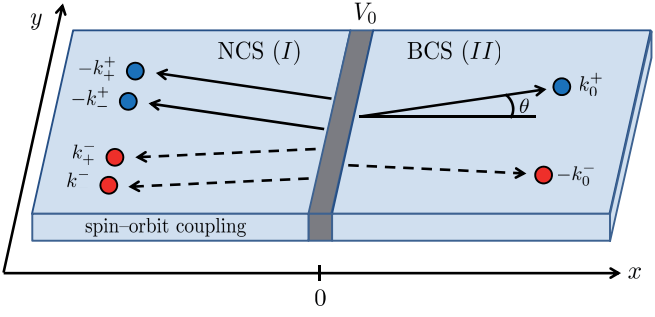}
\caption{Geometry of the 2D Josephson junction. The upper circles indicate electron--like quasiparticles and the lower circles hole--like quasiparticles. Both are labled with the associated wave vectors.\label{cartoon}}
\end{figure}

We consider a Josephson junction in a two--dimensional geometry, as shown in Fig.~\ref{cartoon}, between a non--centrosymmetric superconductor (NCS) on the left hand side ($x<0$) and a conventional $s$--wave BCS superconductor on the right hand side ($x>0$), indicated by I and II, respectively. 
The NCS is described by an intrinsic ASOC of the form $h_{\rm ASOC}=\alpha \vg_\vk\cdot\vtau$ parameterized by a coupling constant $\alpha$, the vector function $\vg_\vk$ describing the anisotropy in momentum $\vk$ and the Pauli matrices $\vtau$. In view of CePt$_3$Si, we choose a Rashba--type of spin--orbit vector $\vg_\vk=(k_y,-k_x,0)^T$~\cite{Frigeri:2004:01,Samokhin:2007:01}.
At the interface, we allow for a spin independent potential barrier $V_0(x)=V_0 \delta(x)$. Due to breaking of the inversion symmetry at the interface~\cite{Gorkov:2001:01,Hayashi:2008:01}, we additionally include a Rashba spin--orbit coupling at the interface
\begin{align}
 h_{\rm SO} &= \alpha_R (\vk\times\vn)\cdot\vtau\, \delta(x) \label{Eq:HRSO}
\end{align} 
where $\vn=(1,0,0)^T$ is the normal vector of the interface. For our restricted two--dimensional geometry this simplifies to
\begin{align}
 h_{\rm SO} &= -\alpha_R \tau_z k_y \delta(x).
\end{align}
Note that this additional term vanishes, if one restricts to perpendicular tunneling.
The order parameter on the conventional superconductor side
\begin{align}
 \vDelta_{\sigma\sigma^\prime}^{II} &= \{\Delta_0 e^{i\phi} i \tau^y\}_{\sigma\sigma^\prime}
\end{align}
is represented by an amplitude $\Delta_0$ and a phase $\phi$.
On side I the order parameter is a superposition of a even--parity spin--singlet ($\psi$) and an odd--parity spin--triplet ($d\vg_\vk$) component:
\begin{align}
 \vDelta_{\sigma\sigma^\prime}^{I}(\vk) &= \{[\psi \vone + d\vg_\vk \cdot \vtau ] i \tau^y\}_{\sigma\sigma^\prime} .
\end{align}
Here $\psi$ and $d$ are assumed to be real and are also used to define the ratio $p=d/\psi$ of the two components. Following Ref.~\cite{Frigeri:2004:02,Samokhin:2008:01} we assume that the $\vd$--vector of the odd--parity component in the NCS is parallel to $\vg_\vk$, as other odd--parity states are suppressed by the ASOC. 

It is convenient to split the Bogoliubov--de--Gennes Hamiltonian in $\vk$--space into three parts corresponding to the three regions of the device, the left hand side ($\hat{\mathcal{H}}^I$), the right hand side ($\hat{\mathcal{H}}^{II}$), and the barrier in the middle ($\hat{\mathcal{H}}^V$):
\begin{align}
H_{\rm BdG} &= \Theta(-x) \hat{\mathcal{H}}^I + \Theta(x)\hat{\mathcal{H}}^{II} +\delta(x)\hat{\mathcal{H}}^V
\end{align}
with
\begin{widetext}
\begin{align}
 \hat{\mathcal{H}}^I &= \left( \begin{array}{cccc}
  \xi_\vk+\alpha g_z & \alpha(g_x-ig_y) & d(-g_x+ig_y) & \psi+dg_z \\
  \alpha(g_x+ig_y) & \xi_\vk-\alpha g_z & -\psi +dg_z & d(g_x+ig_y) \\
  d(-g_x-ig_y) & -\psi+dg_z & -\xi_\vk+\alpha g_z & \alpha(g_x+ig_y) \\
  \psi+dg_z & d(g_x-ig_y) & \alpha(g_x-ig_y) & -\xi_\vk-\alpha g_z
 \end{array} \right) \label{eq:H:NCS:3D} \\
 \hat{\mathcal{H}}^{II} &= \left( \begin{array}{cccc}
  \xi_\vk & 0 & 0 & \Delta_0 e^{i\phi} \\
  0 & \xi_\vk & -\Delta_0 e^{i\phi} & 0 \\
  0 & -\Delta_0 e^{-i\phi} & -\xi_\vk & 0 \\
  \Delta_0 e^{-i\phi} & 0 & 0 & -\xi_\vk
 \end{array} \right) \\
\hat{\mathcal{H}}^{V} &= \left( \begin{array}{cccc}
  \xi_\vk+V_0-\alpha_R k_y & 0 & 0 & 0 \\
  0 & \xi_\vk+V_0+\alpha_R k_y & 0 & 0 \\
  0 & 0 & -\xi_\vk-V_0-\alpha_R k_y & 0 \\
  0 & 0 & 0 & -\xi_\vk-V_0+\alpha_R k_y
 \end{array} \right)
\end{align}
with the abbreviation for the kinetic term
\begin{align}
\xi_\vk &= \frac{\hbar^2\vk^{2}}{2m}-\mu .
\end{align}
Defining the angle $\theta$ of a quasiparticle trajectory  through $k_x=|\vk|\cos\theta$, $k_y=|\vk|\sin\theta$, the Hamiltonian on the NCS side I of the junction reads:
\begin{align}
 \hat{\mathcal{H}}^I &= \left( \begin{array}{cccc}
   \xi_\vk & i\alpha |\vg_\vk| e^{-i\theta} &  -i d |\vg_\vk| e^{-i\theta} & \psi \\
   -i\alpha |\vg_\vk| e^{i\theta} & \xi_\vk & -\psi &  -i d |\vg_\vk| e^{i\theta} \\
   i d |\vg_\vk| e^{i\theta} & -\psi & -\xi_\vk &  -i\alpha |\vg_\vk| e^{i\theta} \\
   \psi &  i d |\vg_\vk| e^{-i\theta} &  i\alpha |\vg_\vk| e^{-i\theta} & -\xi_\vk
 \end{array} \right) \label{eq:H:NCS:3D:Rashba}
\end{align}

The energy eigenvalues on the left and right hand side of the interface are given by $E=\pm E_\lambda$ with $E_\lambda=\sqrt{(\xi_\vk+\lambda\alpha|\vg_\vk|)^2+ \Delta_\lambda^{2}}$ with the index $ \lambda $ being $ \lambda = 0 $ for the BCS superconductor and $ \lambda = \pm 1 $ for the NCS due to the spin--split bands. Accordingly also the pair potential is defined by $\Delta_0$ and $\Delta_\pm=\psi\pm d|\vg_\vk|$ for the two sides. 
%
The corresponding eigenvectors lead to the following ansatz for the wave function on side I (NCS)
\begin{align}
\Psi_{I}(x,y) &=  a_+ \left(\begin{array}{c} u_+ \\ i u_+ e^{-i\theta} \\ -i v_+ e^{-i\theta} \\ v_+  \end{array}\right)
                      e^{-i k^+_{x+} x} e^{i k^+_{y+} y} + 
                  a_- \left(\begin{array}{c} u_- \\ -i u_- e^{-i\theta} \\ i v_- e^{-i\theta} \\ v_- \end{array}\right)
                      e^{-i k^+_{x-} x} e^{i k^+_{y-} y} \nn \\
              &+  b_+ \left(\begin{array}{c} v_+ \\ -i v_+ e^{i\theta} \\ i u_+ e^{i\theta}  \\ u_+ \end{array}\right)
                      e^{i k^-_{x+} x} e^{i k^-_{y+} y} +
                  b_- \left(\begin{array}{c} v_- \\ iv_- e^{i\theta} \\ -i u_- e^{i\theta} \\ u_- \end{array}\right)
                      e^{i k^-_{x-} x} e^{i k^-_{y-} y} \; ,
\end{align}
which is similar to the ansatz used in Ref.~\onlinecite{Asano:2011:01}.
On the side II the ansatz for the wave function reads accordingly as usual for a conventional superconductor, 
\begin{align}
\Psi_{II}(x,y) &=  c \left(\begin{array}{c}  u_0 \\ 0 \\ 0 \\ v_0 e^{-i\phi}  \end{array}\right) e^{ik_{x0}^+ x} e^{i k^+_{y0} y} + d \left(\begin{array}{c} 0 \\ u_0 \\ - v_0 e^{-i\phi} \\ 0  \end{array}\right) e^{ik_{x0}^+ x} e^{i k^+_{y0} y} \\
 &+ e  \left(\begin{array}{c} v_0 \\ 0 \\ 0 \\ u_0 e^{-i\phi} \end{array}\right) e^{-ik_{x0}^- x} e^{i k^-_{y0} y} + f \left(\begin{array}{c} 0 \\ v_0 \\ -u_0 e^{-i\phi} \\ 0 \end{array}\right)  e^{-ik_{x0}^- x} e^{i k^-_{y0} y}
\end{align}
\end{widetext}
with the following wave vectors
\begin{align}
 \hbar k^\pm_{x\lambda} &= \hbar k^\pm_{\lambda} \cos\theta \\
                        &= \sqrt{2m\left( E_F \pm \sqrt{E^2-|\Delta_\lambda|^2}-\lambda\alpha\kF \right)}  \cos\theta \nn \\
 \hbar k^\pm_{y\lambda} &= \hbar k^\pm_{\lambda} \sin\theta \\
                        &= \sqrt{2m\left( E_F \pm \sqrt{E^2-|\Delta_\lambda|^2}-\lambda\alpha\kF \right)}  \sin\theta \nn
\end{align}
and the coherence factors
\begin{align}
 u_\lambda &= \sqrt{\frac{E+\sqrt{E^2-\Delta_\lambda^2}}{2E}} \\
 v_\lambda &= \sqrt{\frac{E-\sqrt{E^2-\Delta_\lambda^2}}{2E}}\frac{\Delta_\lambda}{|\Delta_\lambda|}\;,
\end{align}
with $\lambda=0$ (I) and $\lambda=\pm1$ (II).
%
%

The above ansatz contains the following approximations. In order to obey the Bogoliubov--de--Gennes equations, the wave vectors $k^\pm_{x\pm}$ and $k^\pm_{y\pm}$ have to be expanded to the first non--vanishing order. That is, we replace the wave vectors by $\kF$ in the off--diagonal entries of $\hat H^I$. Whereas we use the full expression on the diagonal for the kinetic energy: $\xi_\vk \rightarrow \hbar^2(k^\pm_{\lambda})^2/2m - \EF = \pm\sqrt{E^2-\Delta^2_\lambda}-\lambda\alpha\kF$. Taking only the leading order in the expansion of the wave vector into account, our equations and thus our results do not depend on the strength of the ASOC $\alpha$. However, the indirect impact of the ASOC through the parity mixing of the order parameter is the dominant effect.
%

The matching conditions for the wave functions at the interface require
\begin{align}
 \Psi_{I}(0,y) &=\Psi_{II}(0,y) \;, \label{boundary:01}
\end{align}
and
\begin{align}
 &\left.\partial_x \Psi(x,y)\right|_{x=0^+} - \left.\partial_x \Psi(x,y)\right|_{x=0^-} = \label{boundary:02} \\
 &\kF\cos\theta \left( \begin{array}{cccc}
  Z^- & 0 & 0 & 0 \\
  0 & Z^+ & 0 & 0 \\
  0 & 0 & Z^+ & 0 \\
  0 & 0 & 0 & Z^-
 \end{array} \right) \Psi(0,y) \; , \nn
\end{align}
taking the barrier potential into account in the latter condition, and including also the spin--orbit coupling term in the interface.  
%
%
It is convenient to define the following dimensionless parameters describing the interface: 
\begin{align}
 Z^\pm &= Z^\prime \pm Z_R \tan\theta =  \frac{2m(V_0\pm\alpha_R \hbar \kF\sin\theta)}{\hbar^2 \kF\cos\theta} \; ,  \\
 Z^\prime &= \frac{Z}{\cos\theta} \; ,\ \ \ Z=\frac{2mV_0}{\hbar^2 \kF} \; , \label{Eq:renormalisation:Z} \\
 Z_R &= \frac{2m\alpha_R}{\hbar\kF} = \frac{\alpha_R p_F}{\EF} \; .
\end{align}
From these boundary conditions it is obvious that the Andreev bound states, emerging at the interface as a subgap part of the spectrum, do not depend on the angle $\theta$ for $Z=Z_R=0$ and are in this sense dispersionless with respect to $ k_y $. On the other hand, for finite scattering potentials $Z$, the spectrum becomes $ \theta $ dependent through the renormalization of the barrier and the back scattering of quasiparticles. This has also an important influence on the Andreev bound states as a function of the Josephson phase $\phi$, as will be discussed later.

In order to have a guide for the barrier parameters, let us here give an estimate for the spin--orbit coupling in the interface. 
For this purpose we assume that the spin splitting of the energy bands at an interface may reach $2|\alpha_R| p_F=0.02 \ldots 0.1\;$eV following Ref.~\onlinecite{Gorkov:2001:01}. Assuming a typical value for the Fermi energy of about $\EF\sim 5\;$eV, we estimate that realistic values for $|Z_R|$ will not exceed $|Z_R| = 0.005 \ldots 0.05$.

\begin{figure}[htbp]
\includegraphics[angle=0, width=0.9\linewidth]{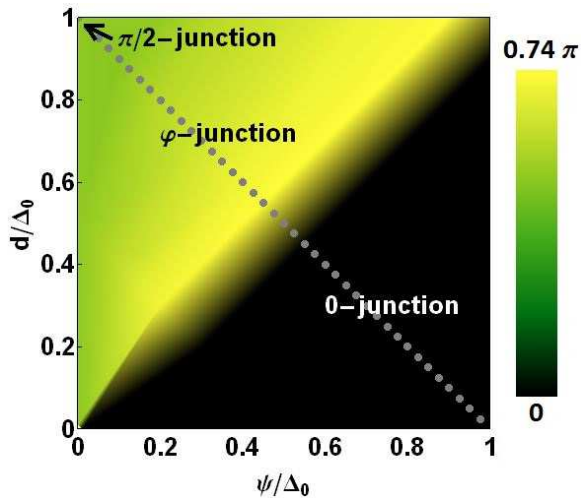}
\caption{(color online) Interpolated phase diagram of the NCS/BCS junction at $Z=0.5$, $Z_{R}=0$, characterized according to the minimum of the junction free energy. Other values of $Z$ and $Z_{R}$ show qualitatively similar results. The dotted line indicates the $\psi+d=\Delta_{0}$ line. }
\label{fig:phasediag:01}
\end{figure}

%
\section{Analytical results\label{analytical:results}}

The Andreev bound states are calculated from Eqs.~(\ref{boundary:01}) and (\ref{boundary:02}), leading up to 8 independent equations. 
Solving this system of linear equations requires that the determinant of the corresponding $8\times 8$ matrix is zero. This yields an implicit and rather lengthy equation for the energy $E$ for given phase difference $\phi$ and angle $\theta$ ($ k_y $). We first consider several limiting cases where a simple analytical expression of the energy can be given, and compare with known results and the numerical calculations in the next section.


As a first limiting case we examine a quasi one--dimensional interface (i.e. $\theta=0$) without Rashba spin--orbit coupling at the interface ($Z_R=0$).
Here, the Andreev bound states are determined easily by the following equation (from here on, we put $\Delta_0=1$ as the energy unit):
\begin{align}
 &8 E^4-8 \left(\Delta _-+\Delta _+\right) E^2 \cos (\phi ) \nn \\
 &+ \Delta _- \Delta _+ \left[8 \cos ^2(\phi )-\Omega^2_0 Z^2 \left(Z^2+4\right)\right] \nn \\
 &+ 4\Omega_0 \left(Z^2+2\right) \left[\Omega_+ \left(E^2-\Delta _- \cos(\phi )\right) \right. \nn \\
 &\left.+ \Omega_- \left(E^2-\Delta _+ \cos (\phi )\right)\right] \nn \\
 &+ \Omega^2_0 \Omega_+\Omega_- \left[\left(Z^2+4\right) Z^2+8\right]  \nn \\
 &+ \Omega^2_0 E^2 Z^2\left( Z^2+4\right) = 0
\label{Andreev_1D}
\end{align}
with $\Omega_\lambda=\sqrt{E^2-\Delta^2_\lambda}$ and $\lambda=\pm,0$.
This expression reproduces the results for the Andreev bound states in asymmetric Josephson junctions as, for example, given, in Ref.~\onlinecite{Wu:2004:01}.

One can further consider a transparent junction (i.e. $Z=0$ and $Z_R=0$), where
the equation which determines the Andreev bound states simplifies considerably:
\begin{align}
 \left( E^2 +\Omega_0\Omega_+ -\Delta_+\cos\phi \right) \\
 \times \left( E^2 +\Omega_0\Omega_- -\Delta_-\cos\phi \right) =0 . \nn 
\end{align}
Since this equation factorizes, we obtain two pairs of solutions for $\Delta_+$ and $\Delta_-$ which can be easily calculated:
\begin{align}
 E^2_\pm (\phi) &= \frac{\Delta^2_\pm \sin^2 \phi}{1+\Delta^2_\pm -2\Delta_\pm \cos\phi} \nn \\
 \mbox{ for } &\ \Delta_\pm \left( 1-\Delta_\pm \cos\phi \right)\left( \Delta_\pm - \cos\phi \right) >0 \; . \label{Eq:analytical:solution:cond}
\end{align}
By rescaling $|\vg_\vk|=1$, the Andreev bound states energies in the singlet limit $\psi=1$, $d=0$, $\Delta_{\pm}=1$ have the solution 
\begin{eqnarray}
E_{\pm}^{2}=\cos^{2}\frac{\phi}{2}
\end{eqnarray}
This is consistent with the $Z=0$ curves in Fig.~\ref{fig:2Dcurrent:10}(b). If instead one considers an interface with large tunneling potential $Z\gg 1$, then Eq.~(\ref{Andreev_1D}) yields
\begin{eqnarray}
E^{2}=\Delta_{-}\Delta_{+}-\Omega_{-}\Omega_{+}\;.
\end{eqnarray}
In the singlet limit this yields $E^{2}=1$. This is consistent with the trend that the bound state energies are flattened with increasing $Z$ in Fig.~\ref{fig:2Dcurrent:10}(b). On the other hand, the transparent junction $Z=0$ at the triplet limit $\psi=0$, $d=1$, $\Delta_{\pm}=\pm 1$ yields 
\begin{eqnarray}
E_{+}^{2}=\cos^{2}\frac{\phi}{2}\;,
\nonumber \\
E_{-}^{2}=\sin^{2}\frac{\phi}{2}\;,
\end{eqnarray}
which are consistent with the $Z=0$ curves in Fig.~\ref{fig:2Dcurrent:00}(b). 
For a ``mixed--parity'' junction, such as the $\psi=0.3$ and $d=0.7$ case studied in Fig.~\ref{fig:2Dcurrent:03}, some branches of Andreev bound states appear only for a limited range of $\phi$ due to the condition in Eq.~(\ref{Eq:analytical:solution:cond}), meaning the subgap states merge with the quasiparticle continuum at the limiting values of $ \phi $.
The condition in Eq.~(\ref{Eq:analytical:solution:cond}) simplifies to $\Delta_- >\cos\phi$ for the special case $\psi +d=1$ (dotted line in Fig.~\ref{fig:phasediag:01}), which will be considered in the following section.

\begin{figure}[htbp]
\includegraphics[angle=0, width=1.0\linewidth]{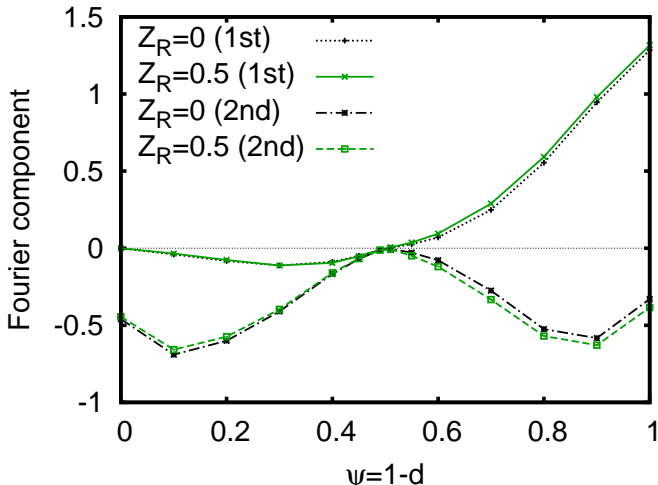}
\caption{First ($\propto\sin(\phi)$, solid and dotted lines) and second ($\propto\sin(2\phi)$, dashed and dash--dotted lines) Fourier component of the total current along the dotted line in the phase diagram of Fig.~\ref{fig:phasediag:01} ($\psi+d=1$) for $Z=0.5$. The dotted and dash--dotted line correspond to $Z_R=0$ and the solid and dashed line to $Z_R=0.5$.\label{fig:fourier:01}}
\end{figure}

\section{Numerical results\label{numerical:results}}
Since for a general set of parameters $\left\{\psi,d,\theta,Z,Z_{R}\right\}$ it is usually not possible to provide an analytical expression for the Andreev bound states energies in our junction, we turn here to numerical solutions. 
As mentioned in the introduction, our interest is to examine the transition in the Josephson tunneling between a pure even and a pure odd--parity superconductor on the NCS side. The parameter space for the gap values which we will explore is illustrated in Fig.~\ref{fig:phasediag:01}.
In this diagram, constant ratios between even and odd--parity component are represented by straight lines through origin. It turns out that the results for the current--phase relation along these lines of constant ratio are at least qualitatively similar. Therefore, we show only some representative examples mainly along the line $\psi+d=1$, labeled by the dots in Fig.~\ref{fig:phasediag:01} and shown in detail in Figs.~\ref{fig:2Dcurrent:10}--\ref{fig:2Dcurrent:00}.
In the following we will characterize the Josephson junction through the phase $ \phi $ at which the junction energy is minimized. The standard junctions have this at phase $ \phi = 0 $ (mod $2 \pi $) and are correspondingly called ``$0$--junctions''. There are also ``$\pi/2$--junctions'' which have minima at $ \phi = \pm \pi/2 $. All other junctions we refer to as ``$\phi$--junctions''. These different junctions have implications on interference experiments using, for example, the superconducting quantum interference devices (SQUID) type of arrangements. 


\subsection{Josephson current}
It has been demonstrated by Chang and Bagwell~\cite{Chang:1994:01} that the Josephson current for an asymmetric contact junction can be decomposed into two contributions. The first one is carried by discrete Andreev bound states, while the second contribution can be assigned to the continuous quasiparticle spectrum above the gaps and becomes important only for strongly asymmetric junctions~\cite{Chang:1994:01}.
Since we are mainly interested in the parity change of the order parameter with $\psi+d=\Delta_0$, we ignore, for simplicity, the continuum contribution to the current, which is expected to give small corrections of the same symmetry as the Josephson current originating from the Andreev bound states at the interface~\cite{Chang:1994:01,Wu:2004:01}.
The expression for the current per unit surface area flowing perpendicular to the surface is then given by Ref.~\onlinecite{Loefwander:2001:01}:
\begin{align}
 j_x &= \frac{e}{\hbar}\sum\limits_{a}\sum\limits_{k_y}\frac{\partial E_a(\phi)}{\partial\phi} f(E_a) \; , \label{eq:current:2D}
\end{align}
where each $E_a(\phi)$ denotes one of the up to four branches in the spectrum of bound states and $f(E)=1/(\exp(E/\kB T)+1)$ is the Fermi--Dirac distribution function. To simplify matters, we restrict to $T=0$. Note that the sum over $k_y$ can be easily converted into an integral over the scattering angle $\theta$.
Furthermore, it is important to account for the right multiplicity of $E_a(\phi)$. The numerical differentiation is then performed after the bound states are assigned to one of the up to four branches.
Eventually, a Fourier analysis of the Josephson current is performed and the free energy $F$ is calculated.

\subsection{Discussion\label{discussion}}

As mentioned above we do not take the continuum states into account, as we can  expect from the discussion in Ref.~\onlinecite{Wu:2004:01} that their effect is limited to an antisymmetric contribution which does not affect the essential conclusions of the current--phase relation.  We will include the spin--orbit coupling of the interface by a rather larger value of $ Z_R = 0.5 $ to compare with the case $ Z_R = 0 $. However, the effect of the spin--orbit coupling is rather weak for our junction geometry.
The general reason for this is the fact that due to the renormalization of $Z$ according to Eq.~(\ref{Eq:renormalisation:Z}), the main contribution to the Josephson current is given by the Andreev bound states for $\theta=0$ where there is no contribution from Rashba spin--orbit coupling according to Eq.~(\ref{Eq:HRSO}).

\begin{figure}[!t]
 \begin{overpic}[angle=0, width=0.85\linewidth]{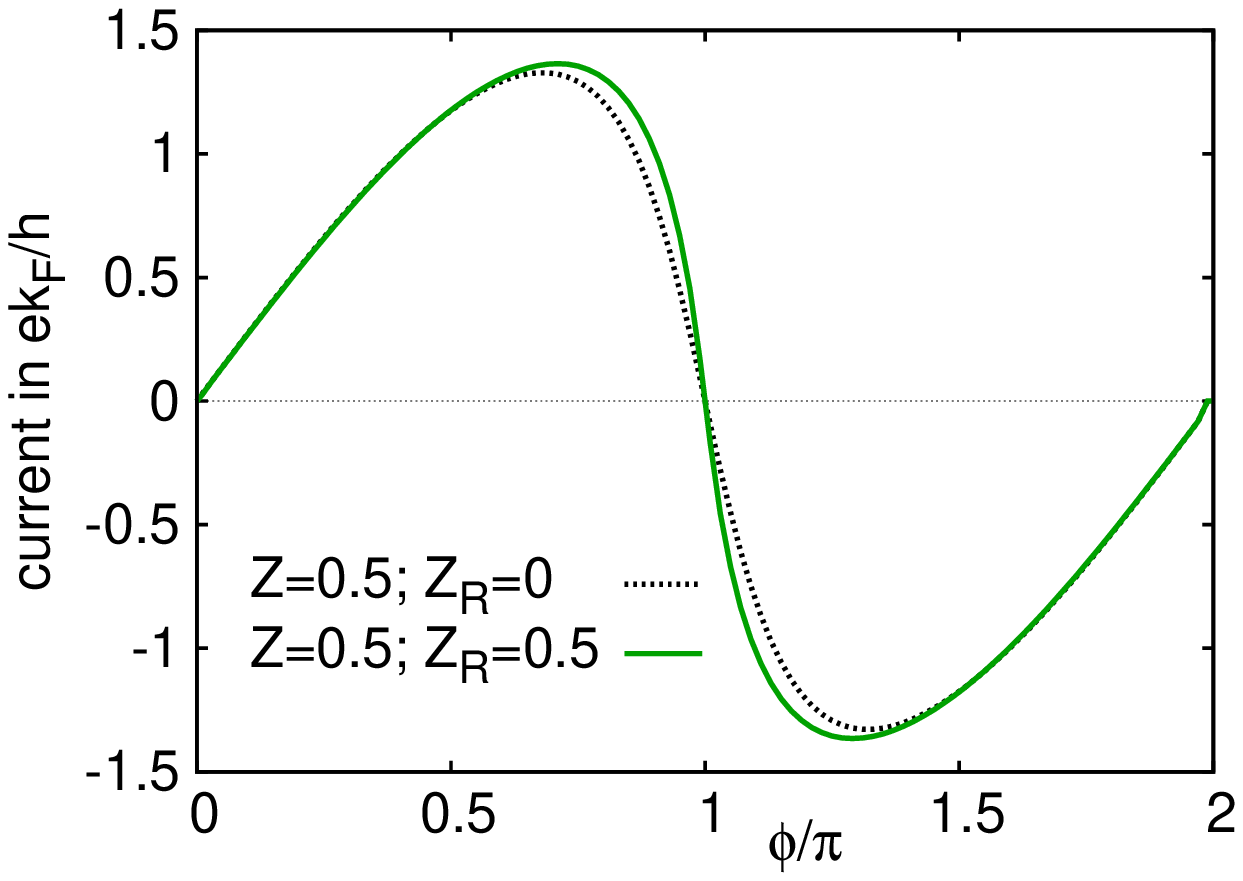}
  \put(58,40.5){\includegraphics[angle=0, width=0.32\linewidth]{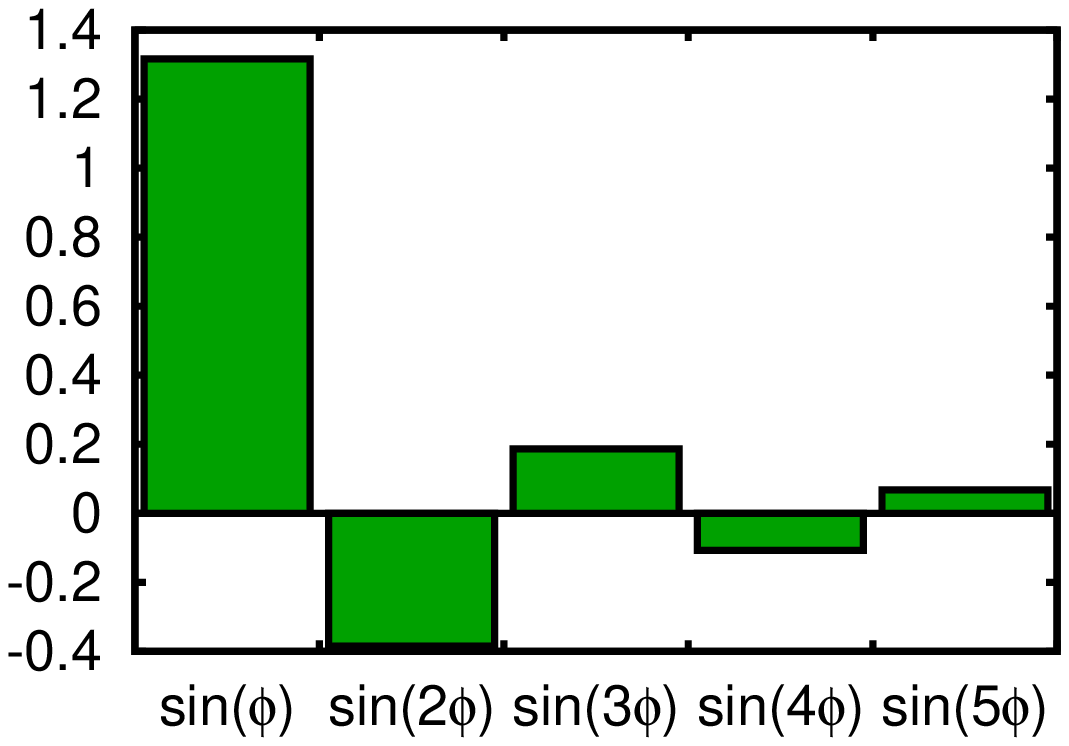}}
  \put(0,65){(a)}
 \end{overpic} \\[0pt]
 \begin{overpic}[angle=0, width=0.85\linewidth]{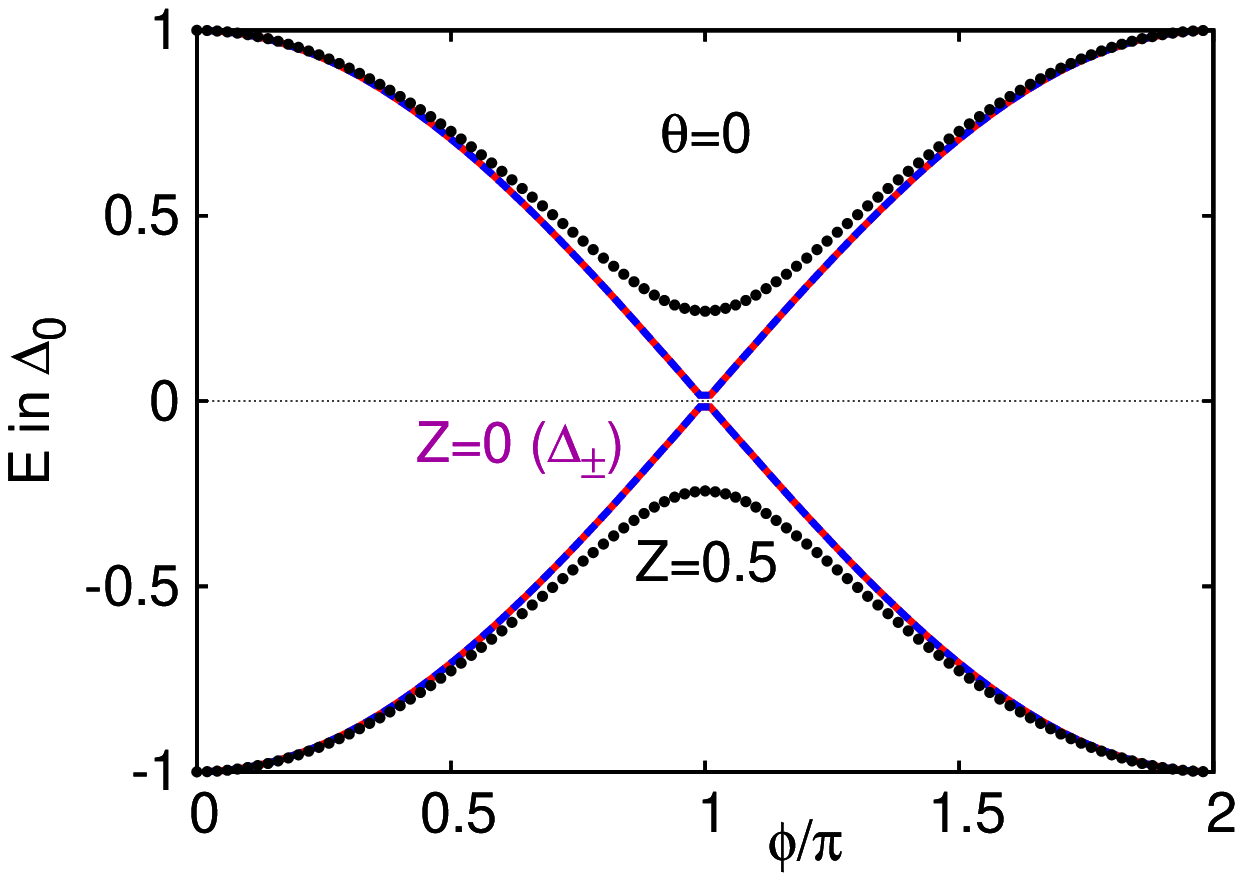}
  \put(0,65){(b)}
 \end{overpic}
\caption{Calculated current--phase relation from the Andreev bound states for a two dimensional $0$--junction with $\psi=1$, $d=0$, $\Delta_0=1$, and $Z=0.5$. The Josephson current (a) is shown for $Z_R=0$ (dotted line) and $Z_R=0.5$ (solid line). Because of the small differences between both curves, the Fourier coefficients are only shown for the $Z_R=0.5$ case as inset. The Andreev bound states (b) are shown for the quasiparticle angle $\theta=0$, once with $Z=0$ (solid line) and once with $Z=0.5$ (filled symbols). Note that the Rashba spin--orbit coupling is idle in this $\theta=0$ case.\label{fig:2Dcurrent:10}}
\end{figure}

\begin{figure}[!t]
 \begin{overpic}[angle=0, width=0.85\linewidth]{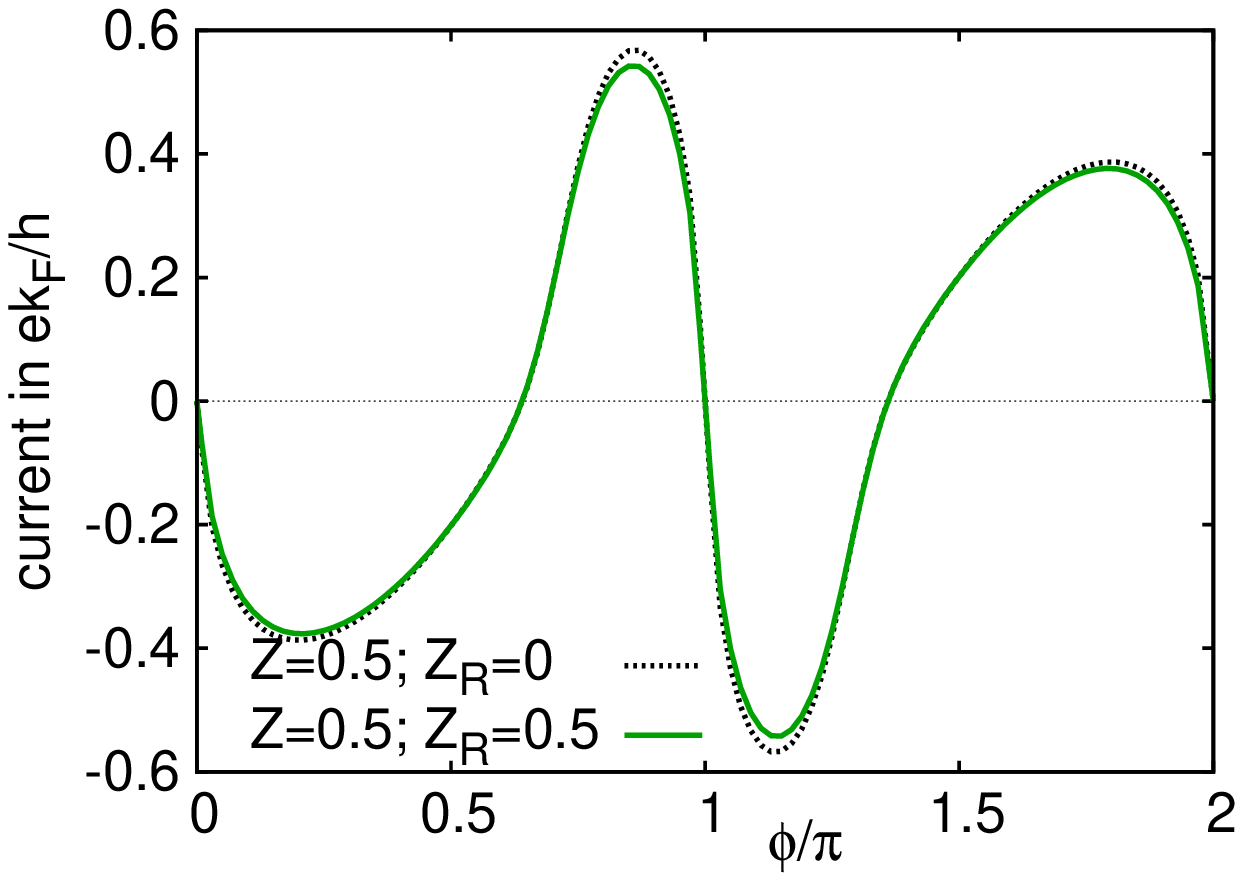}
  \put(58,40.5){\includegraphics[angle=0, width=0.32\linewidth]{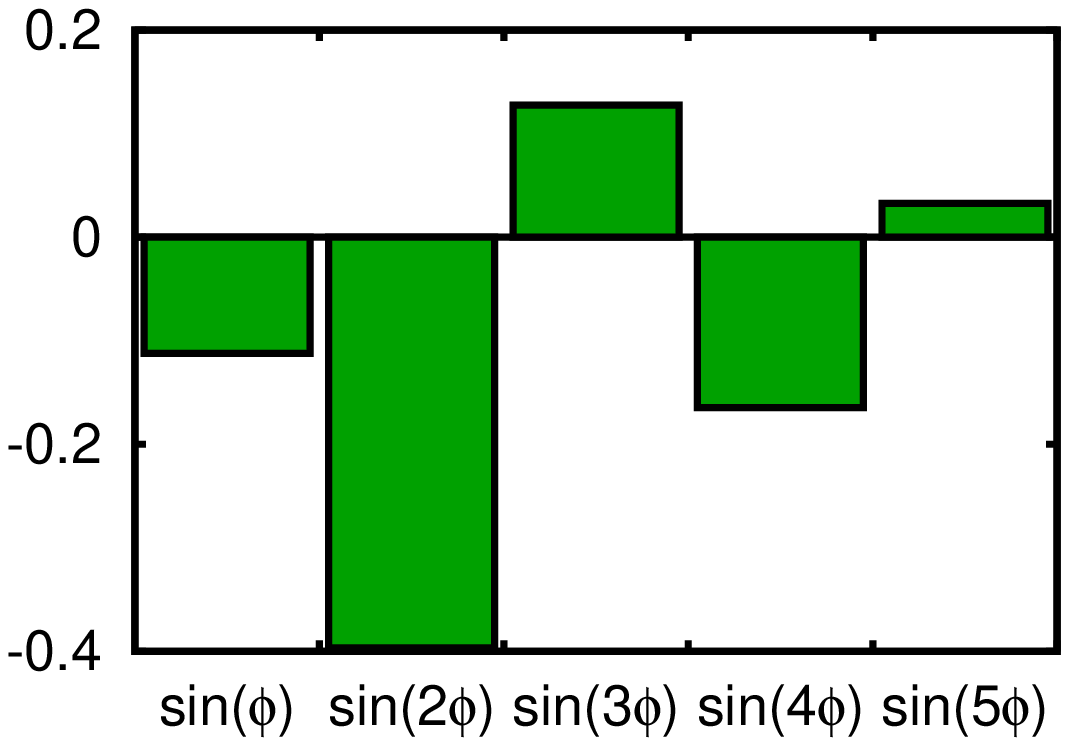}}
  \put(0,65){(a)}
 \end{overpic} \\[0pt]
 \begin{overpic}[angle=0, width=0.85\linewidth]{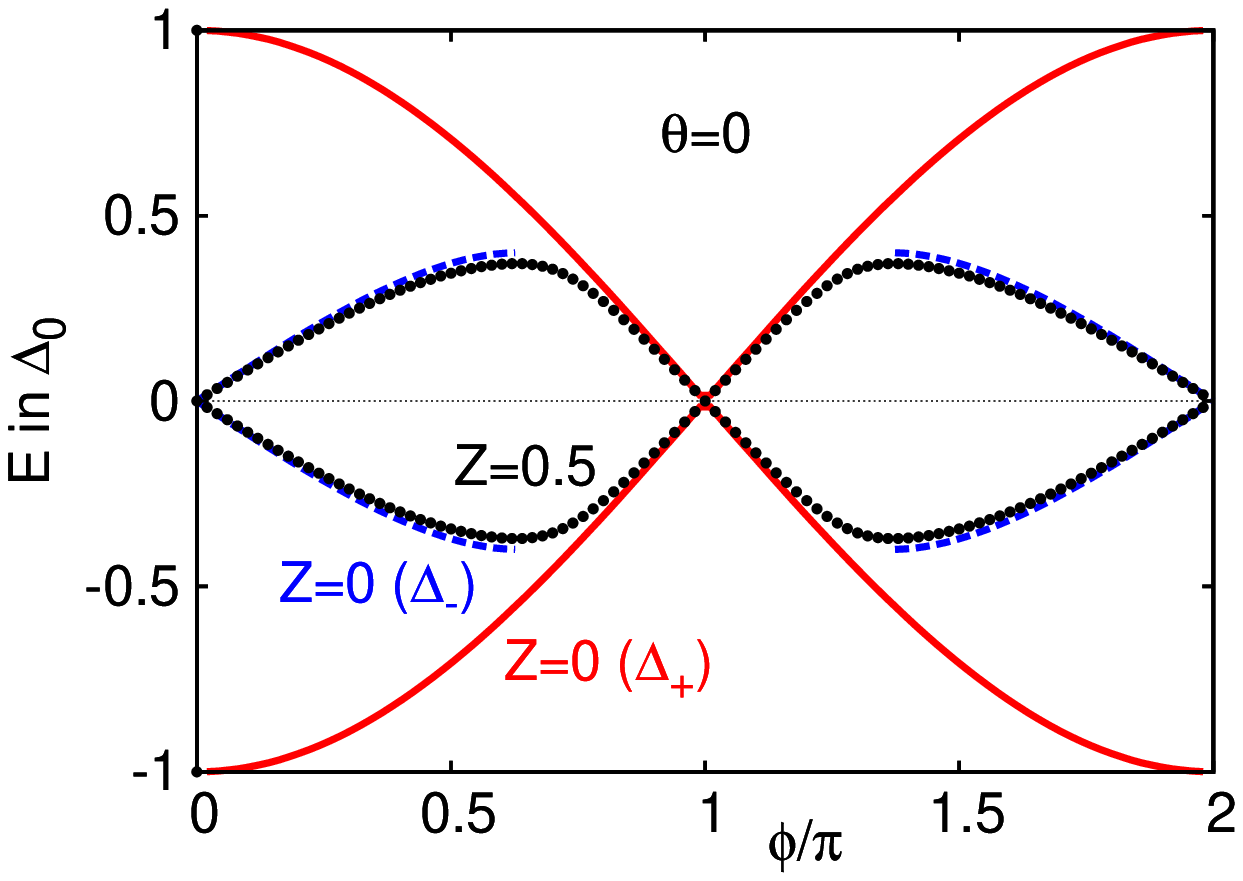}
  \put(0,65){(b)}
 \end{overpic}
\caption{Josephson current (a) for a $\phi$--junction with parameters $\psi=0.3$, $d=0.7$, $\Delta_0=1$, and $Z=0.5$. The minima of the free energy are located at $\phi=0.64\pi$ and $\phi=1.36\pi$. The spin--degenerate Andreev bound states for $\theta=0$ and $Z=0$ are shown in solid and dashed lines, corresponding to the ``$+$'' and ``$-$''--band, respectively. Apart from that, same labelling as in Fig.~\ref{fig:2Dcurrent:10}.\label{fig:2Dcurrent:03}}
\end{figure}
\begin{figure}[!t]
 \begin{overpic}[angle=0, width=0.85\linewidth]{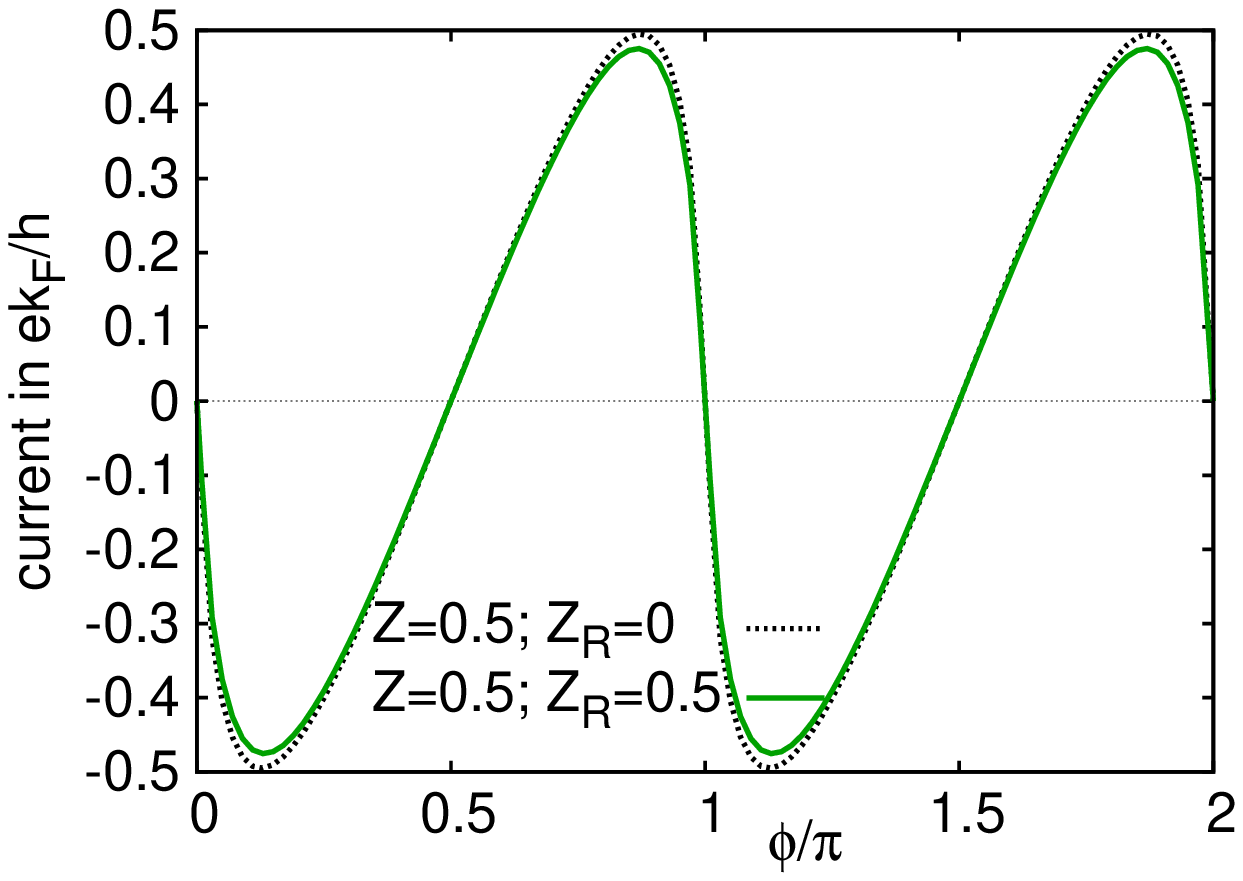}
  \put(58,40.5){\includegraphics[angle=0, width=0.32\linewidth]{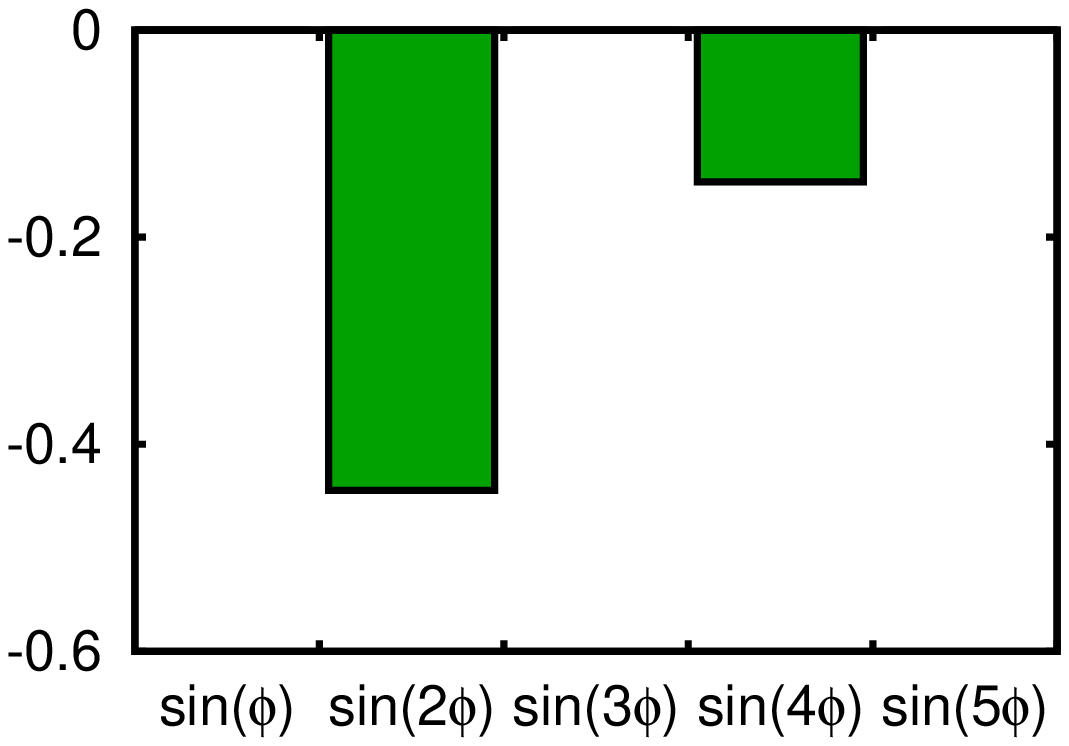}}
  \put(0,65){(a)}
 \end{overpic} \\[0pt]
 \begin{overpic}[angle=0, width=0.85\linewidth]{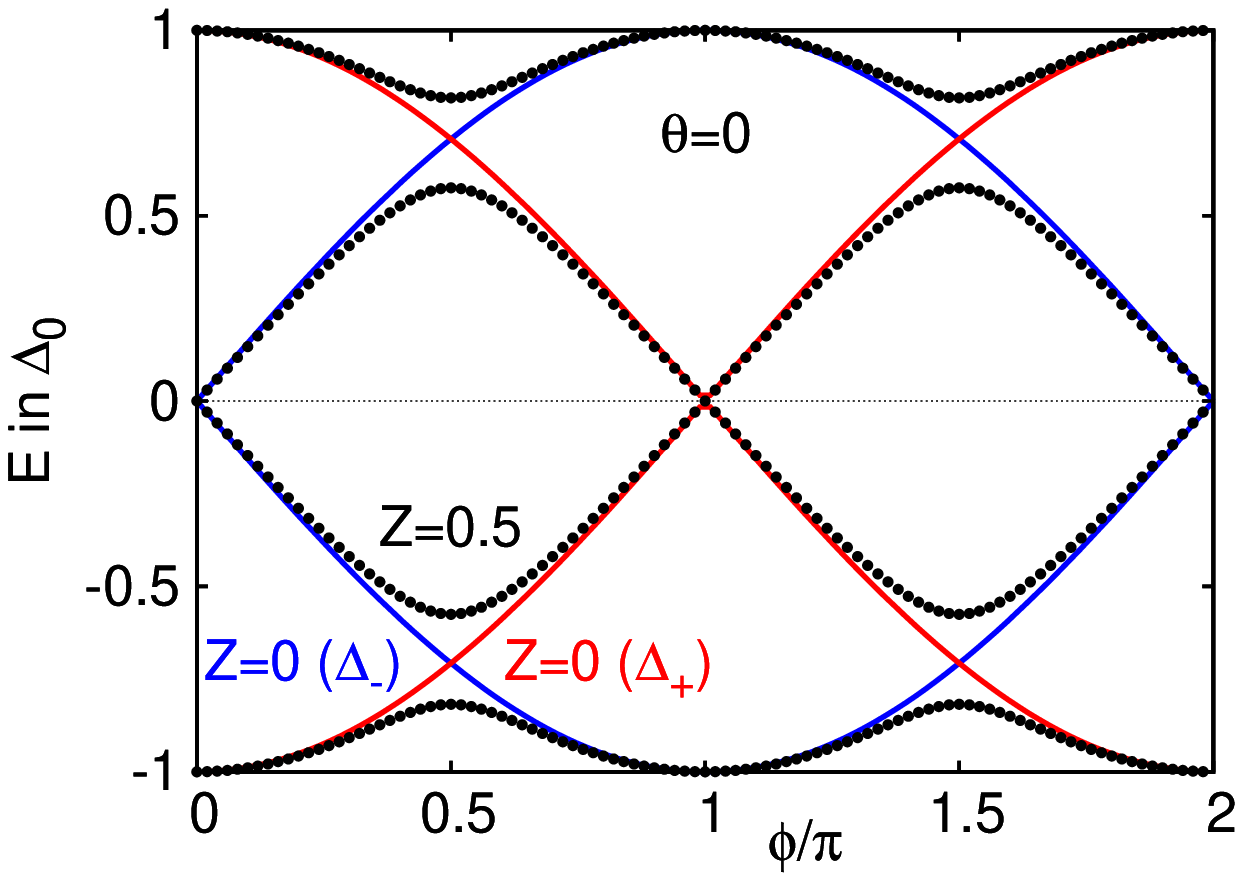}
  \put(0,65){(b)}
 \end{overpic}
\caption{Josephson current (a) for a $\pi/2$--junction with parameters $\psi=0$, $d=1$, $\Delta_0=1$, and $Z=0.5$. Same labelling as in Fig.~\ref{fig:2Dcurrent:03}.\label{fig:2Dcurrent:00}}
\end{figure}

 Now, let us consider the combined results for the Andreev bound states and Josephson current along the line $\psi+d=1$. In order to discuss all relevant and qualitative different cases, we show here results for the parameters $(\psi=1, d=0)$, $(\psi=0.3, d=0.7)$ and $(\psi=0, d=1)$ in Figs.~\ref{fig:2Dcurrent:10}--\ref{fig:2Dcurrent:00}. The upper panel displays the numerical results of the Josephson current for $(Z=0.5, Z_R=0)$ in dotted lines and $(Z=0.5, Z_R=0.5)$ in solid lines as to compare the influence of the interface Rashba spin--orbit coupling. The first five Fourier coefficients of the current--phase relation are shown as an inset with non--zero Rashba spin--orbit coupling. The first two Fourier coefficients with and without Rashba spin--orbit coupling are displayed in Fig.~\ref{fig:fourier:01}, including more data points along the line $\psi+d=1$.
 
In panel (b) of Figs.~\ref{fig:2Dcurrent:10}--\ref{fig:2Dcurrent:00}, the corresponding numerically calculated Andreev bound states are shown in filled symbols. Since we focus on bound states that provide the main contribution to the current, i.e. with $\theta=0$, the Rashba spin--orbit splitting effect on the bound states cannot be seen, as explained above. Therefore it is important to note, that the difference in both displayed current--phase relations (with $Z_R=0$ and $Z_R=0.5$) originates from the bound states with $\theta\neq 0$ which generally give smaller contributions to the Josephson current.
For comparison, we show in panel (b) the analytically calculated spectrum of Andreev bound states from Eq.~(\ref{Eq:analytical:solution:cond}) for a transparent junction ($Z=0$). Turning on the potential scattering ($Z$) at the interface, introduces an anti--crossing between the branch of the bound states associated with $\Delta_+$ (solid line) and the one with $\Delta_-$ (dashed line). This means, that the clear distinction between the branches of bound states belonging to $\Delta_+$ and $\Delta_-$ breaks down.
Since we chose $\psi+d=1$ in Figs.~\ref{fig:2Dcurrent:10}--\ref{fig:2Dcurrent:00}, the order parameter $\Delta_+$ stays constant for all ratios $ \psi/d $, leading to the branch of BCS--like Andreev bound states in the transparent case ($Z=0$) that is identical in all these figures (solid line).

 In Fig.~\ref{fig:2Dcurrent:10} the well--known BCS Josephson junction with $(\psi=1, d=0)$ is recovered as a limiting case. Again, only a small difference can be seen for the current--phase relation with and without Rashba spin--orbit coupling at the interface, shown in panel (a). For this BCS case and  $\theta=0$, the Andreev bound states (panel (b)) are spin--degenerate for $Z=0$ (solid line) as well as for $Z=0.5$ (filled symbols).
This spin--degeneracy is lifted with increasing odd--parity contribution to the gap, as seen, for example, in Fig.~\ref{fig:2Dcurrent:03}(b) for $Z=0$ and $Z=0.5$. The different Andreev bound state spectra for these two cases are due to the anti--crossing at $\phi=0$ and further restrictions due to the finite $Z$.
The special case $\psi=d=1/2$ (not shown in detail here) yields zero--energy bound states due to the node of $\Delta_-$ pointing towards the interface. In this case, the non--transparent junction carries no current from the Andreev bound states. This can be also seen in Fig.~\ref{fig:fourier:01}, where all Fourier components go through zero for $\psi=d=1/2$.
Fig.~\ref{fig:2Dcurrent:03} shows with the parameters $(\psi=0.3, d=0.7)$ a different behavior. The current--phase relation in panel (a) has a negative slope at $\phi=0$, the second Fourier component becomes dominant, and associated with this, the ground state energy of the junction is located at $\phi=0.64\pi$ and $\phi=1.36\pi$. Note that for this case only two particle--hole symmetric and non--degenerate branches of Andreev bound states exist for $Z=0.5$, a feature originating again from the restriction to a finite $Z$.
Finally, the case of $(\psi=0, d=1)$ is displayed in Fig.~\ref{fig:2Dcurrent:00} for the sake of completeness. A detailed discussion of this junction can be found in Ref.~\onlinecite{Lu:2009:01}.

 In the following, we summarize our findings on the transition from a pure BCS Josephson junction to a triplet--singlet junction (Fig.~\ref{fig:fourier:01} and Figs.~\ref{fig:2Dcurrent:10}--\ref{fig:2Dcurrent:00}):

\begin{enumerate}[i)]
\item Despite our relatively large choice for the value of $Z_R$ compared to our estimations, the current--phase relation shows only small differences between $Z_R =0$ and $Z_R =0.5$. Yet the largest differences can be seen for a dominant singlet contribution ($\psi > d$). This is also reflected in the Fourier components of the current, see Fig.~\ref{fig:fourier:01}.
Altogether, the Rashba spin--orbit coupling at the interface leads only to a slightly larger contribution in higher order. Thus, our results are robust with respect to this scattering contribution.

\item From the BCS junction ($\psi=1$, $d=0$) to the triplet--singlet junction ($\psi=0$, $d=1$) the total amplitude of the current shows the following non--monotonous variation. The amplitude is largest for $\psi=1$ and decreases to zero at $\psi=d=1/2$ (at least for finite potential scattering $Z$). For $\psi<1/2$ the amplitude increases again and reaches, after passing over an intermediate maximum, half of the largest value at $\psi=0$. Consequently a weak Josephson coupling (compared to normal state junction resistance) may be an indication for $\psi \approx d$.  

\item From $\psi=1$ to $\psi=0$ the period of the Josephson current changes from $2\pi$ to $\pi$. This results from the vanishing first Fourier component in Fig.~\ref{fig:fourier:01}, while the second component remains finite. 

\item The first Fourier component indicates a transition in the junction through its sign change at $\psi=d=1/2$. 
Clearly, the first Fourier component is dominant for $\psi=1$, it decreases and changes sign at $\psi=d=1/2$ and vanishes again at $\psi=0$ (see Fig.~\ref{fig:fourier:01}). Consequently, the Fourier decomposition of the current--phase relation might be used for an identification of the ratio $ \psi/d $. 

\item The transition at $\psi=d=1/2$ is also indicated by the minimum of the free energy which defines the ground state of the junction. For $\psi>d$ the junction has only one minimum~\footnote{This minima might be extended for some regions in the parameter space. If the contribution from continuum states to the current according to Ref.~\cite{Wu:2004:01} is taken into account, we expect that only one minimum will remain.} around $\phi=0$. For $\psi<d$ the previous minimum at $\phi=0$ becomes a maximum and we obtain two new degenerate ground states located symmetrically around $\phi=\pi$. The position of these two minima are in the proximity of $\phi=\pi$ if $d$ is close to $\psi$, and they shift continuously to the positions $\phi=\pi/2$ and $\phi=3\pi/2$ for ($\psi=0$, $d=1$).
\end{enumerate}



\section{Conclusion\label{summary}}

We investigated the effect of interface Andreev bound states on the Josephson current between a non--centrosymmetric and a conventional $s$--wave superconductor. Using the tetragonal point group symmetry C$_{4v}$ with Rashba--type spin--orbit coupling for the NCS we aim at conditions potentially applicable to non--centrosymmetric heavy Fermion superconductors such as 
CePt$_3$Si, which are especially interesting as candidates for mixed--parity pairing of comparable even and odd--parity components. We apply a general Bogoliubov--de--Gennes formalism and present analytical as well as numerical results for the Andreev bound states and the current--phase relation. Our results show that the behavior of the Josephson current--phase relation is dominated by the tunneling perpendicular to the interface. In this way spin--orbit scattering effects of the interface play  a minor role for the geometry considered (normal vector in the basal plane of the tetragonal crystal lattice). 

We neglect in our analysis the contributions from the continuum of quasiparticle spectrum restricting to the Andreev bound states. Nevertheless, the main result of our study should be unaffected by this constraint. Looking at a changing ratio of even and odd--parity component on the NCS side of the junction, we find that the Josephson current--phase relation changes its character, in a way as to shift the minimum of the junction energy away from $ \phi =0 $ for a conventional BCS--BCS junction 
(0--junction) to finite $ \phi $--values after a transition at a ratio $d/\psi=1$ leading to a $\phi$--junction. The position of the energy minimizing $\phi$ would be one way to figure out the ratio of the parity--mixing state involved. Note that a $\phi$--junction incorporated into a SQUID configuration would generally yield interference pattern distinguishable from standard ones. Another way to determine this unknown ratio is the direct measurement of the current--phase relation, as explained in Ref.~\onlinecite{Golubov:2004:01}. This would give an important tool to characterise the pairing state in a non--centrosymmetric superconductor.

\begin{acknowledgments}
A.~E. thanks the Max Planck Center for financial support and hospitality. L.~K. and M.~S. gratefully acknowledge the financial support from Swiss Nationalfonds.
\end{acknowledgments}

\bibliography{../Bib/CePt3Si_theory,../Bib/CePt3Si_exp,../Bib/not_CePt3Si_exp,../Bib/mypapers,../Bib/miscellaneous,../Bib/coll_modes,../Bib/josephson}

\end{document}